\documentclass[longauth]{aa}

\usepackage{graphicx}
\usepackage{txfonts}
\usepackage{natbib}
\bibpunct{(}{)}{;}{a}{}{,} % to follow the A&A style

\usepackage[small]{caption}

\begin{document}

\title{Observation of VHE $\gamma$-rays from Cassiopeia A with the MAGIC telescope}

\titlerunning{Observation of VHE $\gamma$-rays from Cassiopeia A with the MAGIC telescope}

%\thesaurus{09.19.2; 09.09.1 Cas\,A; 13.07.2; 09.03.2}
\author{ J.~Albert\inst{1}\and E.~Aliu\inst{2}\and H.~Anderhub\inst{3}\and
  P.~Antoranz\inst{4}\and A.~Armada\inst{2}\and
  C.~Baixeras\inst{5}\and J.~A.~Barrio\inst{4} H.~Bartko\inst{6}\and
  D.~Bastieri\inst{7}\and J.~K.~Becker\inst{8}
% ...
}
  \institute {Universit\"at W\"urzburg, D-97074 W\"urzburg, Germany
  \and Institut de F\'\i sica d'Altes Energies, Edifici Cn., E-08193
  Bellaterra (Barcelona), Spain \and ETH Zurich, CH-8093 Switzerland
  \and Universidad Complutense, E-28040 Madrid, Spain \and Universitat
  Aut\`onoma de Barcelona, E-08193 Bellaterra, Spain \and
  Max-Planck-Institut f\"ur Physik, D-80805 M\"unchen, Germany \and
  Universit\`a di Padova and INFN, I-35131 Padova, Italy \and
  Universit\"at Dortmund, D-44227 Dortmund, Germany
% ...
}

\authorrunning{Albert et al}

\date{Received  / Accepted }

\offprints{\\E.~O\~na-Wilhelmi,
\email{emma@apc.univ-paris7.fr\\V.~Vitale \email{vitale@fisica.uniud.it}}}

\abstract{}{We searched for very high energy (VHE) $\gamma$-ray emission from the
  supernova remnant Cassiopeia A}{The shell-type supernova remnant
  Cassiopeia A was observed with the 17 meter MAGIC telescope between
  July 2006 and January 2007 for a total time of 47~hours.} {The
  source was detected above an energy of 250 GeV with a significance
  of 5.2~$\sigma$ and a photon flux above 1~TeV of
  (7.3$\pm$0.7$_{stat}\pm$2.2$_{sys})\times10^{-13}
  cm^{-2}s^{-1}$. The photon spectrum is compatible with a power law
  dN/dE $\propto$E$^{-\Gamma}$ with a photon index
  $\Gamma$=2.3$\pm$0.2$_{stat}$$\pm$0.2$_{sys}$. The source is
  point-like within the angular resolution of the telescope.} {}


\maketitle

\section{Introduction}

Cassiopeia A (Cas~A), with right ascension (RA) and declination (DEC)
(23.385$^h$,58.800$^o$), is a prominent shell type supernova remnant
and a bright source of synchrotron radiation observed at radio
frequencies, see \cite{Bell}; \cite{Tuffs}, and in the X-ray band, see
\cite{Allen}, \cite{Favata}. The remnant results from the youngest known Galactic supernova, whose explosion took place around 1680. Its distance was estimated at 3.4 kpc by \cite{Reed}. High resolution X-ray images from the Chandra
satellite, see \cite{Hughes}, reveal a shell-type nature of the
remnant and the existence of a central object. The progenitor of Cas~A
was probably a Wolf-Rayet star, as discussed in \cite{Fesen} and
\cite{Iyudin}. The progenitor's initial mass was large, estimated to be between 15 and 25~M$_{\odot}$, see \cite{Young}. The morphology of the remnant as seen in optical, X-ray and IR wavelength consists on a patchy and irregular shell with a diameter of 4' (4~pc at 3.4~kpc). The supernova blast wave is expanding into a wind bubble formed from the previous wind phases of the progenitor star; this plays an important role in shock acceleration of CR, see \cite{Berezhko}.

At TeV energies, Cas~A was detected by the HEGRA Stereoscopic
Cherenkov Telescope System, which accumulated 232 hours of data from
1997 to 1999. TeV $\gamma$-ray emission was detected at 5$\sigma$
level and a flux of
(5.8~$\pm$~1.2$_{stat}\pm$1.2$_{sys}$)~10$^{-13}$~ph~cm$^{-2}$s$^{-1}$
above 1~TeV was derived, as discussed in \cite{HEGRA}. The spectral
distribution between 1 and 10 TeV was found to be consistent with a
power law with a differential spectral index of
-2.5$\pm$0.4$_{stat}\pm$0.1$_{sys}$. Upper limits at TeV energies have
been set also by Whipple, see \cite{Lessard} and CAT, see
\cite{Goret}. These upper limits were consistent with
the HEGRA detected flux level. At lower energy, EGRET set an upper
limit for a flux below $12.4\times10^{-8}$~cm$^{-2}$s$^{-1}$, see \cite{EGRET}.

The HEGRA detection makes Cas~A a good scenario to test the supernova
remnant emission at lower energies, in particular for trying to
further distinguish between leptonic and hadronic models for the
origin of the $\gamma$-ray emission. A summary of the observations and
analysis results is given in Section \ref{Observations}, the results
are reported in Section \ref{Results} and finally a comparison of
MAGIC detection with the existing model predictions for the TeV
$\gamma$-ray emission on Cas~A is discussed in Section \ref{Summary}.

%%%%%%%%%%%%%%%%%%%%%%%%%%%%%%%%%%%%%%%%%%%%%%
\section{Observations and Data Analysis}     %
%%%%%%%%%%%%%%%%%%%%%%%%%%%%%%%%%%%%%%%%%%%%%%

\label{Observations}

The MAGIC (Major Atmospheric Gamma Imaging Cherenkov) Telescope is located on
the Canary Island La Palma (2200~m asl, $28^\circ45' N$, $17^\circ54'W$) and
has a 17 m-diameter tessellated reflector dish, see ~\cite{Cortina}. The total
field of view is 3.5$^\circ$. The accessible energy range spans from 50-60~GeV
(trigger threshold at low zenith angle) up to tens of TeV. The telescope
angular resolution (sigma of the Gaussian fit to the point spread function,
PSF, $\sigma_{psf}$) is about 0.09$^\circ$.

Cas~A observations were performed between June 2006 and January 2007 for a
total observation time of 47~hours after quality cuts, namely, after rejecting
runs with detector problems or adverse atmospheric conditions. The zenith
angle ranged from 29$^{\circ}$ to 45$^{\circ}$ and averaged 35$^{\circ}$. The
observation technique applied was the so-called wobble mode, see \cite{Daum}
in which the telescope pointed alternatively for 20 minutes to two opposite
sky positions at 0.4$^{\circ}$ off the source.
Most of the data were taken under moderate moonlight illumination
(86$\%$ of the scheduled observation time). Depending on the different
moonlight levels, the resulting PMT anode currents ranged between
1~$\mu$A and 6~$\mu$A, as compared to a typical anode current of
1~$\mu$A for dark night observations. Correspondingly, the trigger
discriminator threshold (DT) was varied between 15 and 30 arbitrary
units (a.u.) to keep a low rate of accidental events. The mean trigger
discriminator threshold during the observations was 19 a.u, which
corresponds to 13.3~photoelectrons (PE). Briefly, the impact of the
rise of DT can be summarized by a decrease on the relative
$\gamma$-ray efficiency from 1 (dark observations) to 0.84 while the
relative sensitivity\footnote{\it Minimal flux detectable with
$5\sigma$ significance in 50~hours of observations.} worsens from 2.5
to 2.7$\%$ with respect to the Crab flux. Although this effect is
important for images containing a low number of PE's (low size), the
energy threshold rise ( $\sim$ 5~GeV) is negligible compared to the
rise due to the medium to high zenith angle. Hence, the moderate
moonlight illumination did not substantially reduce the telescope
performance. More details on the Moon data analysis are discussed by
\cite{Moon}.
\begin{figure}[!tp]
\centering
\includegraphics[width=10cm]{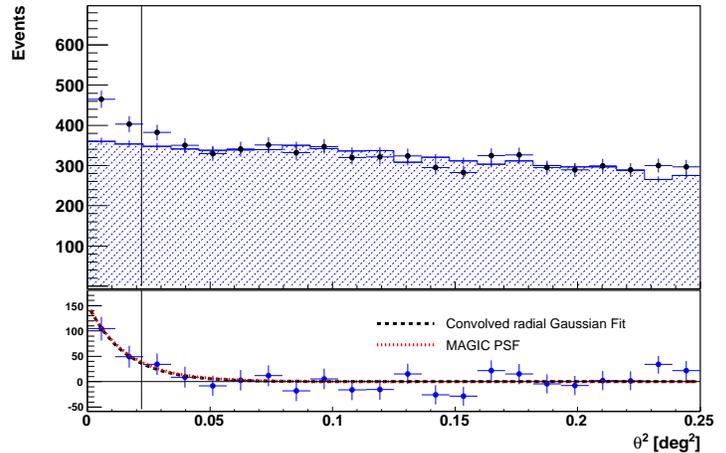}
\caption{The upper panel shows the distributions of $\theta^2$ 
    (measured in degree$^2$) in the direction of the source (black
    dots) and anti-sources (blue shaded histogram). A lower SIZE cut
    of 400~PE was applied. The lower panel shows the $\theta^2$
    distribution after background subtraction. The vertical line shows
    the optimum angular cut. The red function corresponds to the
    telescope PSF and is derived from Crab nebula data taken in the
    same observational conditions as those for Cas~A. The black
    distribution is the result of a Gaussian fit to the excess
    distribution.}
\label{theta2}
\end{figure}
Dark and Moon data were analyzed together using the standard analysis
and calibration programs for the MAGIC collaboration,
e.g. \cite{gaug}. The images were cleaned using absolute tail and
boundary cuts of 10 and 5~PE, respectively. For the $\gamma$/hadron
shower separation, the shower images were parameterized using the
Hillas parameters, see \cite{Hillas}. These variables were combined
for $\gamma$/hadron separation by means of a Random Forest
classification algorithm, see~\cite{Bock}, trained with MC simulated
$\gamma$-ray events and data from galactic areas near the source under
study but containing no $\gamma$-ray sources. The Random Forest method
calculates for every event a parameter dubbed HADRONNESS (H), which
parameterizes the purity of hadron-initiated images in the
multi-dimensional space defined by the Hillas variables.

The $\theta^2$ distribution is computed for the source position, where
$\theta$ is the angular distance between the source position in the
sky and the reconstructed origin position of the shower. The
reconstruction of individual $\gamma$-ray arrival directions makes use
of the so-called DISP method (\cite{disp}). The expected number of
background events are calculated using five regions symmetrically
distributed for each wobble position with respect to the center of the
camera and refered to as anti-sources.

The optimum H and the angular cuts were derived using dark night Crab
data of the same epoch and in the same observation conditions (zenith
range, astronomical nights). The use of a dark night data sample in
optimizing the telescope sensitivity is justified by the results in
\cite{Moon}. For the spectral analysis, the energy of each individual
$\gamma$-ray candidate was also estimated using the Random Forest
technique. The average energy resolution for the analyzed energy range
was 20$\%$.

%%%%%%%%%%%%%%%%%%%%%%%%%%%%%%%%%%%%%%%%%%%%%%%%%%%%%%%%%%%%%
\section{Source detection, extension and energy spectrum}   %
\label{Results}                                             %
%%%%%%%%%%%%%%%%%%%%%%%%%%%%%%%%%%%%%%%%%%%%%%%%%%%%%%%%%%%%%

The so-called $\theta^2$ distributions for the source and anti-source
positions are shown in Figure~\ref{theta2} for a lower SIZE cut of
400~PE, which optimizes the MAGIC signal to noise ratio. The black
points correspond to the source position whereas the blue-shaded
histogram corresponds to the anti-sources.  The subtraction of the two
histograms shows the excess in the direction of Cas~A.  An excess of
$N_{excess}$=157 with a significance of 5.2~$\sigma$ (using the
likelihood method of \cite{LiMa}) is detected within the region
0.13$^{\circ}$ centered at the HEGRA position.

\begin{figure}[!tp]
\centering
\includegraphics[width=7cm]{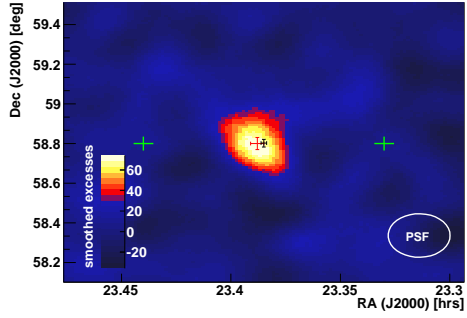}
\caption{Sky map around the position of Cas~A. A lower cut in SIZE of
    400~PE was applied. The green crosses mark the 2 wobble
    positions. The red cross indicates the MAGIC best fit
    position. The black cross marks the HEGRA source position, which
    is within 1 standard deviation from the MAGIC one. The bars of the
    crosses for both the MAGIC and HEGRA marks correspond to 1 sigma
    statistical errors.}
\label{CasAmap}
\end{figure}

Figure~\ref{CasAmap} shows the excess map of $\gamma$-ray candidates with
images larger than 400~PE. The map has been smeared with a Gaussian of
$\sigma$=0.07$^o$. The source position has been determined by ways of a fit of
the non-smeared sky map to a bidimensional Gaussian function. The best fit
position coordinates are RA = 23.386$\pm$0.003$_{stat}\pm$0.001$_{sys}$~h and
DEC = 58.81$\pm$0.03$_{stat}\pm$0.02$_{sys}^{\circ}$ (for more details on the
systematic uncertainties in the source position determination, see
\cite{pointing}).

In X-rays and radio-frequencies Cas~A has an angular diameter of
0.08$^{\circ}$, which is just on the limit of the MAGIC angular
resolution.  The MAGIC system PSF is derived from MC simulation for a
point source, and is found to be
$\sigma_{psf}$=0.090$\pm$0.002$^{\circ}$ (shown in Figure
\ref{CasAmap}). This value was validated with Mkn 421 and Crab Nebula
data (see \cite{crab}). To further constrain the extension of the
source we fit the excess with a Gaussian function convolved with the
PSF (F=$A\cdot exp(-0.5~\theta^2/(\sigma_{src}^2 +
\sigma_{psf}^2$))). We obtain a value of $\sigma_{src}$ which is
compatible with zero within the fit error.  Figure~\ref{theta2} shows
the telescope PSF and the result of the Gaussian fit (dotted blue
curve).

\begin{figure}[!tp]
\centering
\includegraphics[width=9.2cm]{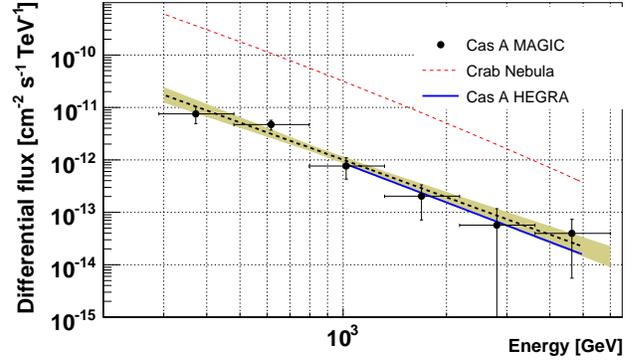}
\caption{Cas~A spectrum above 250 GeV. The blue line represents the earlier
    measurement by HEGRA. The red line represents the Crab nebula
    spectrum. The shaded area is the 1$\sigma$ statistical error of
    the fit.}
\label{Casspectrum}
\end{figure}

Figure~\ref{Casspectrum} shows the reconstructed spectrum above
250~GeV. The spectrum is consistent with a power law (dN/dE $\propto$
E$^{-\Gamma}$). The differential flux at 1~TeV is
(1.0$\pm$0.1$_{stat}\pm$0.3$_{sys})\times10^{-12}$
TeV$^{-1}$cm$^{-1}$s$^{-1}$ with a photon index of
$\Gamma$=2.4$\pm$0.2$_{stat}$$\pm$0.2$_{sys}$.The systematic error is
estimated to be 35$\%$ in the flux level determination and 0.2 in the
spectral index (see \cite{IC443}). The measured spectrum was unfolded
using the Gauss-Newton method, see \cite{unfolding}. The
$\chi^2$/d.o.f of the fit is 2.83/3. The 1~$\sigma$ error limit on the
flux fitted is also added as a grey band. The Cas A flux corresponds
to an integral flux above 1~TeV of 3$\%$ of the Crab nebula flux above
the same threshold (in red dashed line in figure~\ref{Casspectrum},
see \cite{crab}). The Cas A spectrum measured by HEGRA is also shown
as a blue solid line. The spectrum measured about 8 years later by
MAGIC is consistent with that measured by HEGRA for the energies above
1~TeV, i.e, where they overlap.
%
%
%
%%%%%%%%%%%%%%%%%%%%%%%%%%%%%%%%%%%%%%%%%%%%%%%%%%%%%%%%%%%%%
\section{Discussion}                                       %
%%%%%%%%%%%%%%%%%%%%%%%%%%%%%%%%%%%%%%%%%%%%%%%%%%%%%%%%%%%%%
\label{Summary}

The VHE MAGIC 47-hour observation of Cas~A confirms the source
detection by HEGRA after a multi-year integration of 232~hours and at
the same time significantly extends the energy spectrum down to about
250 GeV. Cas~A is detected with more than 5~$\sigma$ at a flux level
compatible with the HEGRA measurement for those energies explored in
common. The diferential flux at 1 TeV measured by MAGIC is
1.0$\pm$0.1$_{stat}\times 10^{-12}TeV^{-1}cm^{-1}s^{-1}$ to be compared
with the one measured by HEGRA, 0.9$\pm$0.2$_{stat}\times
10^{-12}TeV^{-1}cm^{-1}s^{-1}$. The agreement between the two measures
is excellent not only in the determination of the flux level but also
in the spectral index measured. Although the errors in the spectral
index are large, there is no evidence for a high energy cutoff, nor
for a deviation from a power law at lower energies.
The detection of very high energy $\gamma$-rays from Cas~A provides
evidence of the acceleration of multi-TeV particles in SNR shocks and
their visibility in gamma-rays \cite{Drury}.

Significant efforts have been made for the theoretical modeling of
Cas~A's multi-frequency emission, including that at the highest
energies. 
The effect of an energy-dependent propagation of relativistic
electrons in a spatially inhomogeneous medium has been used in order
to interpret the radio emission from the region and define its
electron content (\cite{Atoyana}). The variations in brightness in the
radio band is so complex that a multi-zone model was used: distinguishing
between compact, bright spectrum radio knots and the bright fragmented
radio ring on one hand, and a diffuse plateau on the other. A
three-zone model with a magnetic field decreasing from its highest value
in the compact zones putatively related with acceleration sites, to a
lower value in regions surrounding the shell, to yet a lower value in
the neighborhood has been found to reproduce the radio data, with a
magnetic field around and below 1~mG. The fluxes at TeV energies, due
to Bremsstrahlung and inverse Compton radiation of the same
relativistic electrons have also been computed (\cite{Atoyanb}) and,
albeit the parameters allow a large range of possible fluxes, the
overall shape of the spectrum remains similar, showing a steep cutoff
for multi-TeV energies (see, e.g., Figure 7 of \cite{Atoyanb}). This
cutoff is not seen in HEGRA and/or MAGIC data, disfavoring a leptonic
origin of the radiation.
Vink and Laming (2003) also studied multi-zone models for Cas A,
assuming no difference between zones other than in their magnetic
field. They found that an IC origin of high energy fluxes would be
possible but only for low values (within the range allowed to be
consistent with radio and X-ray observations, see e.g., Vink and
Laming 2003, Hwang et al. 2004) of the magnetic field and high,
far-infrared photon density. The generally high values of the magnetic
field necessary to explain the multi-frequency observations makes it
likely that TeV emission from Cas A is then dominated by pion decay
(Atoyan et al. 2000b, Vink and Laming 2003).

\begin{figure}[!tp]
  \centering
  \includegraphics*[width=0.5\textwidth,angle=0,clip]{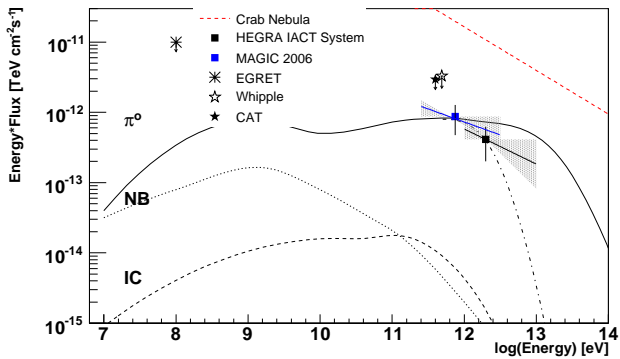}
\caption{\label{Multi} {Spectra of Cas~A as measured by MAGIC. The
    shaded area around the 0.65~TeV detection shows the 1$\sigma$
    statistical error range under the assumption of a E$^{-\alpha}$
    power law spectrum.  The upper limits given by Whipple, EGRET and
    CAT are also indicated, as well as the HEGRA detection. The MAGIC
    and HEGRA spectra are shown in the context of the model by
    \cite{Berezhko}. Both hadronic ($\pi^o$ with and without an energy
    cutoff) and leptonic (NB and IC) $\gamma$-ray emission are
    shown. The normalization of the pion decay spectrum can be taken
    as a free parameter.}}
\end{figure}

\cite{Berezhko} applied a non-linear kinetic model of cosmic-ray acceleration 
to describe Cas~A, ignoring the role of
any small scale inhomogeneities for the production of the very high
energy particles and considering the whole SNR blast wave as the main
relativistic particle generator.
Figure~\ref{Multi} represents the expected integral $\gamma$-ray flux
components from non-thermal Bremstrahlung NB, IC scattering on the
background radiation field (cosmic microwave + optical/infrared), and
hadronic collisions of CR protons with gas nuclei, respectively, for
this model. The pion-decay $\gamma$-ray flux presented in
Figure~\ref{Multi} --with and without an exponential cutoff at 4 TeV--
was calculated with a renormalization factor of 1/6 (i.e. this factor
takes into account that not all the SNR shock efficiently injects and
accelerates cosmic rays). This emphasizes that the normalization of
nucleonic predictions of $\gamma$-rays is to be considered a free
parameter, within certain reasonable boundaries.
The predicted slope for the dominating nucleonic-produced
$\gamma$-rays (that dominates, even when all possible uncertainties
leading to an increase of the leptonic emission are included) is hard
in the range of interest, as shown in Figure~\ref{Multi}, perhaps too
hard already to provide a good fit to the new MAGIC data at low
energies. Higher and lower energy measurements, and a better signal to
noise ratio for the spectrum determination of such a weak source, are
still needed for a definite answer.

\begin{acknowledgements}
  We would like to thank the IAC for the excellent working conditions at the
  Observatorio del Roque de los Muchachos in La Palma.  The support of the
  German BMBF and MPG, the Italian INFN and the Spanish CICYT is gratefully
  acknowledged. This work was also supported by ETH Research Grant TH~34/04~3
  and the Polish MNiI Grant 1P03D01028.
\end{acknowledgements}

\bibliographystyle{apj}
\bibliography{casamag.bbl}

\listofobjects
\end{document}